**Natural and fishing mortalities affecting eastern sea garfish, *Hyporhamphus australis***

**inferred from age-frequency data using hazard functions**

Matt K. Broadhurst[a,b]*, Marco Kienzle[c,d], John Stewart[e]

[a]*NSW Department of Primary Industries, Fisheries Conservation Technology Unit, National Marine Science Centre, PO Box 4321, Coffs Harbour, NSW 2450, Australia*

[b]*Marine and Estuarine Ecology Unit, School of Biological Sciences, University of Queensland, Brisbane, QLD 4072, Australia*

[c]*Department of Agriculture and Fisheries, Ecosciences Precinct, 41 Boggo Road, Dutton Park, Brisbane, QLD 4072, Australia*

[d]*University of Queensland, School of Agriculture and Food Sciences, St Lucia, QLD 4072, Australia.*

[e]*NSW Department of Primary Industries, Sydney Institute of Marine Science, Chowder Bay Road, Mosman, NSW 2088, Australia*

*Corresponding author. tel: +61 2 6648 3905; email: matt.broadhurst@dpi.nsw.gov.au*

**Abstract**

Estimates of age-specific natural ($M$) and fishing ($F$) mortalities among economically important stocks are required to determine sustainable yields and, ultimately, facilitate effective resource management. Here we used hazard functions to estimate mortality rates for eastern sea garfish, *Hyporhamphus australis*—a pelagic species that forms the basis of an Australian commercial lampara-net fishery. Data describing annual (2004 to 2015) age frequencies (0–1 to 5–6 years), yield, effort (boat-days), and average weights at age were used to fit various stochastic models to estimate mortality rates by maximum likelihood. The model best supported by the data implied: (i) the escape of fish aged 0–1 years increased from approximately 90 to 97% as a result of a mandated increase in stretched mesh opening from 25 to 28 mm; (ii) full selectivity among older age groups; (iii) a constant $M$ of $0.52 \pm 0.06$ year$^{-1}$; and (iv) a decline in $F$ between 2004 and 2015. Recruitment and biomass were estimated to vary, but increased during the sampled period. The results reiterate the utility of hazard functions to estimate and partition mortality rates, and support traditional input controls designed to reduce both accounted and unaccounted $F$.

# 1. Introduction

Hemiramphidae is cosmopolitan teleost family comprising >60 species; most of which are targeted throughout their tropical and temperate distributions by various active and passive fishing gears (Berkeley and Houde, 1978; Sokolovsky and Sokolovskaya, 1999; Jones, 1990; McBride and Thurman, 2003; Stewart and Hughes, 2007). In Australia, >15 species are sought, with the endemic eastern sea garfish, *Hyporhamphus australis* among the most economically important (Stewart and Hughes, 2007).

Considered a unit stock, almost the entire catch of *H. australis* is landed by up to 50 beach- and boat-based commercial fishers using lampara nets off central and southern New South Wales (NSW) (Stewart et al., 2004). Traditionally, *H. australis* fishers used a minimum mesh size of 25 mm (stretched mesh opening; SMO) which along with excessive effort led to declines in annual catches from peaks of ~200 t in the early 1990s to ~35 t after 2007 (present study and Stewart et al., 2004). In response to concerns over stock status and in the absence of detailed population assessments, the minimum SMO in lampara nets was increased to 28 mm (initiated in 2006 but not fully adopted until 2009–10 as fishers replaced their existing nets) and effort was reduced. These input controls resulted in the fishery constantly landing ~45–50 t per annum (Stewart et al., 2015). Ancillary recreational (mostly hook-and-line) catches of *H. australis* are minor (<5 t per annum; Henry and Lyle, 2003); only ~15% of which are discarded, and with variable associated mortality, depending on their handling (Butcher et al., 2010).

Notwithstanding a stabilized harvest of *H. australis*, very few data are published describing their population dynamics (but see Stewart and Hughes, 2007) and despite such information considered an essential prerequisite for effectively managing exploitation (King, 2007). Pivotal among population-dynamic parameters to assess fish stocks (and ultimately harvestable yields) are estimates of age-specific natural ($M$) and accounted fishing ($F$) mortalities which, along with unaccounted fishing mortality, are summed to provide total mortality ($Z$). Estimating mortality rates and the factors affecting the survival of individuals in a population has been the preoccupation of scientists for at least three centuries (Chiang, 1968).

For many fisheries, mortalities are estimated using stock-assessment models. Natural mortality is one of the most important stock-assessment parameters for fishery management, but is notoriously difficult to estimate (Lee et al., 2011). Consequently, many researchers have used empirical relationships between life-history traits and environmental factors to estimate $M$ (Pauly, 1980; Hoenig, 1983; Jensen, 1996; Then et al., 2014). Similarly, when the data required for sophisticated stock assessment models are unavailable (e.g. from fishery-independent surveys), $F$ often is derived simply by subtracting $M$ from $Z$, with the latter estimated using catch curves (King, 2007). Other methods for estimating $Z$ have included standard linear regression on log-transformed age-frequency data, but recent studies imply these should be avoided (e.g. Miller, 2014).

There exist long-term age-frequency data (from otolith readings; Stewart and Hughes, 2007) for *H. australis* which have been used within the above types of empirical relationships by Stewart et al. (2005) to provide preliminary estimates of $Z$ and $M$ at 3.0–3.4 and 0.8, respectively. However, in the absence of robust hypothesis testing, such values are subjected to numerous assumptions and lack the necessary rigor to derive other important population parameters.

As an alternative approach to concurrently estimate $M$ and $F$ using age-frequency data, here we propose statistical methods developed beyond the field of fisheries research. Specifically, in the middle of last century, demographers, epidemiologists, insurers and zoologists developed methods to analyse age data for a cohort (or a group of cohorts) and presented the results in the form of life tables. Many researchers showed that the age distributions of dead individuals can be used to infer mortality rates (e.g. Caughley, 1966; Spinage, 1972). This approach subsequently received formal mathematical treatment, from the point of view of the theory of stochastic processes by Chiang (1968), who formulated the joint probability distribution of the number of deaths, which ultimately provided the statistical foundation to estimate mortality parameters by maximum likelihood. More recently, Cox and Oakes (1984) introduced key concepts and definitions under the banner of 'survival analysis', presenting statistical methods to estimate mortalities for practitioners in areas as diverse as medical research, actuarial sciences, engineering and econometrics.

There are few examples where survival analysis methods have been applied in fisheries (Dupont, 1983; Neilson et al., 1989; Ferrandis, 2007; Kienzle, 2016) or marine science more broadly (Stolen and Barlow, 2003; Manocci et al., 2012). Despite limited application, survival analysis applies readily to cases where age data are measured from a sample of catch. The traditional approach of partitioning mortality as $M$ and $F$ falls naturally into the competing-risk framework, which is a statistical method to construct life tables accounting for multiple sources of mortality (Chiang, 1968; Cox and Oakes, 1983; Dupont, 1983). Further, unlike other stock assessment methods, survival analysis does not require an index of abundance to estimate the dynamics of a fishery: it can be applied when only age samples from catch-and-effort data are available (Kienzle, 2016).

Considering the above, our aims here were to quantify the mortality of *H. australis* through a hazard-function approach. Further, we sought to explore the utility of such analysis as an alternative to traditional stock assessment methods for characterising aspects of the associated dynamics, including estimated recruitment and biomass (population quantities).

## 2. Material and methods

### 2.1. Data

Four data groups were obtained from various sources to estimate the required parameters and quantities describing the entire stock of *H. australis* between 2004 and 2015. The first data group comprised the total number of fish aged in each year, which varied between 216 and 462 otolith measurements collected from commercial catches at several landing sites in NSW (see Stewart and Hughes, 2007 for methodology). In the model section 2.2, these data are denoted as a matrix $S = [s_{i,j}]$, $i = 0,1,\dots,10$ and $j = 0,1,\dots,5$. Index $i$ represents (fishing) year intervals from 2004–5 to 2014–15. Index $j$ represents age-groups from 0–1 to 5–6 (see Table 1 in supplementary material, for details). The data contain 16 cohorts, numbered by convention from the top right-hand side to the bottom left-hand side of the age sample matrix ($S$) using $k$ varying from 0 to 15. The number of age-groups ($r_k$) belonging to each cohort varies between 1 and 6, and is given by:

$$r_k = \begin{cases} k+1 & \text{if } k < 5, \\ 6 & \text{if } 5 \leq k < 11, \\ 16-k & \text{if } 11 \leq k \end{cases}$$

The second data group was annual yields derived from landed catch records from year intervals 2004–5 to 2014–15. The annual yields are defined as $Y = [y_i]$, where index $i$ represents the year interval the same as index $i$ of $s_{ij}$ (see Table 3 in supplementary material, for details). The third data group was annual fishing effort obtained from mandatory logbooks from year intervals 2004–5 to 2014–15. The annual effort is defined as $E = [e_i]$, where index $i$ represents year interval the same as index $i$ of $y_i$ (see Table 3 in supplementary material, for details). The last data group was average weight for each age group caught each year calculated from individual weights collected during surveys of commercial landings. The data are denoted as a matrix $W = [w_{i,j}]$, where indices $i$ and $j$ are the same as $s_{i,j}$ of $S$ (see Table 4 in supplementary material).

Several quantities were derived from these data to perform subsequent calculations. First, the proportion $p_{i,j}$ of fish in age group $j$ observed in year $i$ was calculated using $s_{i,j}$:

$$p_{i,j} = s_{i,j} / \sum_{x=0}^{5} s_{i,x} \tag{1}$$

Second, the estimated annual number of fish $c_i$ caught in year $i$ was calculated using $y_i$, $p_{i,j}$ and $w_{i,j}$:

$$c_i = \frac{y_i}{\sum_{x=0}^{5} p_{i,x} \, w_{i,x}} \tag{2}$$

Finally, the estimated number of fish $\tilde{c}_{i,j}$ of age-group $j$ caught in year $i$:

$$\tilde{c}_{i,j} = c_i p_{i,j} \qquad (3)$$

## 2.2. Stochastic models and mortality estimation

A stochastic population model was used to estimate mortality rates by maximum likelihood using the matrix of age sample ($S$) based on a competing-risk model (Chiang, 1968; Dupont, 1983; Ferrrandis, 2007) assuming two sources of mortality ($M$ and $F$) affected *H. australis*. Natural mortality was represented by an 11-row × six-column matrix ($M$), the elements ($m_{i,j}$) of which give the magnitude of hazard due to natural causes that fishes are exposed to in each year interval. Similarly, $F$ was represented by an 11-row × six-column matrix ($F$) with elements ($f_{i,j}$) quantifying the magnitude of $F$ fishes are exposed to during each year interval. In order to build a life table for each cohort $k$, we calculated the probability of members of each cohort dying from fishing in an interval $j$ given they were alive at the beginning of this interval (Chiang, 1968):

$$\mu_{k,j} = \frac{f_{k,j}}{m_{k,j}+f_{k,j}} \left[ 1 - e^{-(m_{k,j}+f_{k,j})} \right] \qquad (4)$$

In addition, we calculated the probability for individuals in each cohort to survive until the beginning of an interval $j$ as the product of surviving all preceding intervals based on the magnitude of mortalities in each interval (i.e. the cumulative survival in Quinn and Deriso (1999)'s terminology and the survivor function described in the survival analysis literature (Cox, 1983)):

$$\vartheta_{k,j} = e^{-\left[ \sum_{x=j_{min}}^{j} (m_{k,x}+f_{k,x}) - (m_{k,j}+f_{k,j}) \right]} \qquad (5)$$

For $j = \begin{cases} 5-k \leq j \leq 5 & if \ k < 6 \\ 0 \leq j \leq 5 & if \ 5 \leq \ k < 11 \\ 0 \leq j \leq 15-k & if \ 11 \leq \ k \end{cases}$

Therefore, the probability that members of a cohort die from fishing in an interval $j$ is the product of surviving until the beginning of that interval, and the probability of dying from fishing within that interval (Chiang, 1968):

$$\pi_{k,j} = \vartheta_{k,j} \times \mu_{k,j} \tag{6}$$

The equations above define a probability of dying from fishing for each element of the matrix $S$. The data in matrix $S$ did not encompass the entire lifespan of many cohorts represented in this dataset: some had information for a single age group only (e.g. cohort 0 at the top-right corner of the age sample matrix and cohort 15 at the bottom-left) and some cohorts were missing either younger or older age-groups (e.g. cohorts 1–3 and 12–14, respectively). In other words, the probabilities of dying from fishing were not proper (they did not sum to 1), and so the probability distribution functions (PDF) used in the likelihood function for each cohort were truncated. Therefore, the probabilities of dying from fishing were normalized over the range of ages available for each cohort:

$$\tilde{\pi}_{k,j} = \frac{\pi_{k,j}}{\sum_x \pi_{k,x}} \tag{7}$$

Various mortality models, represented by matrices $M$, $F$ and $Z$, where $z_{i,j} = m_{i,j} + f_{i,j}$, were fitted to determine which was best supported by the data. All models assumed the $M$ matrix was constant with all its elements equalled to $m$. A $F$ matrix was constructed according to the separability assumption as the product of (i) gear selectivity ($G$; a matrix with rows representing the selectivity pattern at age); (ii) fishing effort ($\tilde{E}$; a matrix with identical columns equal to $E$); and (iii) a constant catchability matrix ($q$). Subsequently, the $Z$ matrix was expressed as:

$$z_{i,j}(\theta) = m + \underbrace{q\,\tilde{e}_{i,j}\,g_{i,j}}_{=f_{i,j}} \tag{8}$$

Based on known morphology (i.e. maximum girth and TL) and estimates of gear selectivity by Stewart et al. (2004), all models assumed that *H. australis* aged 0–1 were partially selected by the lampara nets $\left(0 \leq g_{i,j} \leq 1\right)$. The remaining ages were assumed to be fully selected $g_{i,j} = 1$, for $\forall i \geq 1$). Models were fitted to the data to estimate parametric characteristics for different hypotheses.

The first model had three parameters $\left(\theta = (m, s, q)\right)$ representing constant (i) $M$, (ii) selectivity $(s)$ of 0–1 age *H. australis* throughout the entire time series; and (iii) catchability $(q)$. The second model had an additional selectivity parameter to capture a change in selectivity for age 0–1 after 2009–10 (when all fishers were assumed to have adopted the 28-mm mesh in their lampara nets). This second model estimated a vector of four parameters $\left(\theta = (m, s_1, s_2, q)\right)$. In the third model, $M$ was fixed at 0.7 year$^{-1}$ following the methods of Hoenig (1983) and Pauly (1980) and using growth-function parameters described in Stewart and Hughes (2007), while $q$ and $s$ were estimated.

Given that each fish in a cohort eventually dies within an age group encompassing the range of ages observed in the population, the joint probability of catch follows a multinomial distribution (Chiang, 1968). The combinatorial term of the joint probability of catch, which is a constant depending only on the data and not any parameters of the mortality model, disappears when the likelihood is differentiated to estimate its parameters (Rodriguez, 2007). The sample of number-at-age (S) was corrected for varying sampling and fishing intensities into a matrix S* because Monte Carlo simulations in Kienzle (2016) showed that parameters estimates are biased when this correction is not applied. The parameters were estimated by maximizing the following likelihood function (Stolen and Barlow, 2003) in R (R Core Team, 2016):

$$\mathcal{L}(\theta) = \prod_k \prod_j \tilde{\pi}_{k,j}{}^{S^*{}_{k,j}} \tag{9}$$

### 2.3. Deriving relevant population quantities

According to Quinn and Deriso (1999), the abundance of fish ($N$) at the beginning of each time interval relates to the matrix of number of individuals caught in each yearly interval $i$ and age-group $j$ ($\tilde{C}$) through $M$ and $F$ affecting that age group during that year according to the following relationships:

$$n_{i,j} = \frac{\tilde{c}_{i,j}}{\mu_{i,j}} \tag{10}$$

$$\mu_{i,j} = \frac{f_{i,j}}{m_{i,j}+f_{i,j}}(1-e^{-(m_{i,j}+f_{i,j})}) \tag{11}$$

The total yield in a year interval ($Y$) is the product of the number caught at age ($\tilde{C}$) and the average weight-at-age of individuals $W$:

$$y_i = \sum_{x=0}^{5} \mu_{i,x}\ n_{i,x}\ w_{i,x} \tag{12}$$

The average biomass of the stock ($B$) in each year is:

$$b_i = \sum_{j=0}^{5} n_{i,j}\ \frac{1-e^{-(m_{i,j}+f_{i,j})}}{m_{i,j}+f_{i,j}} w_{i,j} \tag{13}$$

Recruitment, defined as the number of fish alive at age 0 at the beginning of each year ($R_i$) was estimated following Dupont (1983) using for each cohort $k \geq 5$ as the ratio of the total number of individuals caught and the sum of probability of dying from fishing over all age-groups:

$$R_{k-5} = \frac{\sum_x \tilde{c}_{k,x}}{\sum_x p_{k,x}} \tag{14}$$

Minimization of the logarithm of the likelihood function, $L(\theta)$, was used to determine the best parameter estimates $\tilde{\theta}$ for the vector of parameters $\theta$. A region around the best likelihood estimates that contains the true values of the parameters with a certain probability ($W$) was defined according to Brandt (1998):

$$L(\theta) = L(\tilde{\theta}) + \frac{1}{2}\,\chi^2_W(n_f) \tag{15}$$

where $L(\theta)$ is the value of the log-likelihood function for values $\theta$, $L(\tilde{\theta})$, is the minimum of the log-likelihood function and $\chi^2_W(n_f)$ is the quantile of the $\chi^2$-distribution for $n_f$ degrees of freedom and probability $W$. The 95% confidence interval of biomass and recruitment estimates were calculated using $W = 0.95$ and $n_f = 4$ (Bolker, 2008).

## 3. Results

### 3.1 Temporal changes in catch and effort

Fishing effort steadily declined from 901 fishing boat-days in 2004–5 to a fairly consistent 237–203 boat-days from 2010–11 (Fig. 1). By comparison, excluding an unusually high catch in 2009–10, yield remained at around $37 \pm 7$ tonnes throughout the time series (Fig. 1). As a consequence, nominal catch-per-unit-of-effort more than tripled from around $50 \pm 2$ kg boat-days$^{-1}$ at the beginning of the time series to $186 \pm 28$ kg boat-days$^{-1}$ in the three most recent years.

### 3.2 Model parameter estimates

Of the three models fitted to the data, the second (with an additional parameter to represent a change in selectivity among age 0–1 years after 2009–10) provided the best fit according to the value of the log-likelihood (Table 1). This model estimated that the selectivity on age 0–1 years declined from 10.3 to 3.5% (Table 2).

Fishing mortality among age 0–1 years estimated with model 2 declined through time from 0.17 year$^{-1}$ to 0.01 and from 1.7 to 0.38 for age groups ≥1–2 years (Table 2; Fig. 2). For those ages fully selected by the lampara nets (i.e. ≥1–2 years), $F$ was substantially greater during the first two years of the surveyed period, but then steadily declined to below $M$ (estimated at $0.52 \pm 0.06$ year$^{-1}$ throughout) until 2010–11, when it remained fairly stable at ~$0.40 \pm 0.15$ year$^{-1}$ (Table 2; Fig. 2). Fishing mortality among *H. australis* aged 0–1 years was hypothesized to reflect the change in lampara-net selectivity, varying from $0.17 \pm 0.02$ to $0.1 \pm 0.01$ year$^{-1}$ prior to 2009–10, before dropping to $0.01 \pm 0.001$ after 2010–11 (Table 2; Fig. 2a). When combined, these partitioned mortalities provide estimates of $Z$ for ages 0–1 and >1–2 years of $0.54 \pm 0.06$ and $0.91 \pm 0.16$ year$^{-1}$, respectively for the most recent years.

The estimated temporal recruitment of *H. australis* remained variable (peaking in 2008–09 and then in 2013–14), but in absolute terms increased over the time series from <1.8 million individuals to >2 million (Fig. 3a). Estimated biomass also increased from <150 t before 2009–10 to >250 t in 2014 (Fig. 3b).

## 4. Discussion

This study provides the first robust assessment of partitioned, instantaneous mortalities for the Australian endemic *H. australis* and contributes towards the limited, but quite variable estimates—and especially for $M$—available for Hemiramphidae from other regional (*H. melanochir*: $M$ = ~0.55; Jones, 1990) and overseas congenerics (*H. brasiliensis* and *H. bala*: $M$ = ~ 0.75 to 1.15; Mahmoudi and McBride, 2002). The estimated mortalities, selectivities and subsequent recruitment and biomass for *H. australis* can be discussed with respect to life-history traits, possible extrinsic environmental variables, and the temporal changes in fisheries management. But first, the relative utility and limitations of our chosen analytical method warrant consideration.

An implicit assumption of any modelling approach for estimating population dynamics is adequate vector accuracy. Here, the likelihood function provided reasonable log-likelihood values. During a comparison of multinomial likelihood and likelihood functions from survival analyses, Kienzle (2016) demonstrated that the latter had better performances as sample sizes increased and provided unbiased estimates. Here, annual sample sizes of aged fish ($n$ = 216–462) were relatively small, and could account for some variability in uncertainty among parameter estimates. Similarly, the estimated fishing effort was based on logbook entries with the implicit assumption of accurate reporting, although any biases should be consistent and boat-days might be considered fairly robust compared to finer-scale estimates such as actual lampara-net deployments.

Another consideration is the potential for changes in fishing power that might encompass a positive increase over time with improved technology (Marchal et al., 2006) and/or shifts in fisher behaviour. For example, increasing the SMO from 25 to 28 mm meant that greater numbers of small (with similar girths to the mesh perimeter) *H. australis* were meshed; increasing the time required to retrieve nets and lowering catch prices because of scale loss and damage (Stewart et al., 2004; Butcher et al., 2010). As a consequence, some fishers possibly avoided setting their lampara nets around perceived schools of small *H. australis*. Notwithstanding the above caveats, we consider our estimates to be accurate representations of absolute parameters, and especially their relevant temporal variation (i.e. for $F$).

The hazard-function approach and mortalities estimated here contrast with earlier studies on species within Hemiramphidae, including *H. australis*. Specifically, using empirical methods, Stewart et al. (2005) estimated collective (across all ages) $M$ and $F$ at 0.8 and 2.2–2.6, respectively; both of which are up to 1.5 × larger than the estimates here. Such differences would considerably impact recruitment and biomass assessments and ultimately estimates of sustainable yield. In particular, our estimated $M$ (0.52 ± 0.06 year$^{-1}$) for *H. australis* was within one standard deviation of that estimated for the more southern distributed *H. melanochir* (0.55 ± 0.13 year$^{-1}$; Jones, 1990), but substantially lower than for other heavily exploited Hemiramphidae, including the sub-tropical *H. brasiliensis* and *H. balao* (0.75 to 1.15 year$^{-1}$; Mahmoudi and McBride, 2002). Such differences might imply a latitudinal gradient, which seems coherent considering the potential importance of temperature on mortality (Gislason et al., 2010). But the latter estimates were derived from observed maximum ages as described in Mahmoudi and McBride (2002), rather than formal modelling. Clearly, robust approaches are required to elucidate accurate assessments and facilitate broader relative comparisons among species with possibly divergent life-history strategies.

Notwithstanding uncertainty among absolute estimates of $M$, like for several small schooling pelagics (e.g. *Clupeonella* spp.; Daskalov and Mamedov, 2007) including other Hemiramphidae (Berkeley and Houde, 1978), we noted the potential for a quite large $F$ among *H. australis*—estimated at up to 1.70 (or ~82%) during the first year of available data (2004–05). These mortalities reflect the ease of catchability (with encircling gear like lampara nets) as these species aggregate in schools to feed and/or presumably seek protection from predation (Larsson, 2012).

The dense aggregation of schools and their high catchability also meant that once effort and gear selectivity was appropriately regulated, there was a rapid reduction in $F$ to below $M$ during the last three years, and with a concomitant plateau in catches (not withstanding variable recruitment and biomass). This outcome not only supports the utility of the simple applied regulations, but also in the case of minimum mesh sizes (and assuming not all small *H. australis* were avoided by all fishers), validates the previously untested assumption of relatively low associated escape mortality. Logistics typically preclude accurately estimating escape mortality from mobile gears (discussed by Broadhurst et al., 2006), although implied impacts here might be derived from related work. For example, considering Butcher et al. (2010) demonstrated low discard mortalities among *H. australis* that were angled and then released without being handled (i.e. quickly shaken off the hook), it would not be unreasonable to assume that fish escaping from a slowly hauled lampara net would be minimally impacted.

While the estimated recent $M$ and $F$ are low compared to previous studies and certainly would reduce the potential for recruitment overfishing, the species clearly has susceptibility for high $F$—owing to their schooling behaviour and ease of accessibility. The apparent variable recruitment also compounds any possible population impacts associated with increasing $F$. The susceptibility for such variable recruitment is likely to reflect various extrinsic variables that might include, but are not limited to, El Niño and La Nina and their effects on water temperature, rainfall and subsequent primary production (Gillson, 2011). For example, during the monitored period, two mild La Nina events, with greater than normal rainfall in NSW, occurred during 2007–8 and 2008–9, followed by a stronger La Nina in 2010–12 (and with almost twice the annual precipitation). These periods were subsequently followed by peaks in the recruitment and biomass of *H. australis*. Other extrinsic variables are also likely to be important, and warrant assessment for their influences on recruitment (not only for *H. australis*, but also other similar schooling species) to facilitate appropriate effort control.

The results from this study have several important implications. First, the outcomes reiterate the utility of survival analysis for estimating partitioned natural and fishing mortalities (Dupont, 1983; Ferrandis, 2007; Kienzle, 2016). Second, it is evident that the selected hazard-function model allowed the effects of management changes to be assessed—which included both an increase in mesh size and an almost 80% reduction in fishing effort to a fairly stable 200 boat-days in 2011. Lastly, it is apparent that the management changes were effective in reducing $F$ to a constant level approaching $M$ and promoting what now appears to be a sustainable, small-scale fishery.

**Acknowledgements**

This study was funded by the NSW Department of Primary Industries and the Queensland Department of Agriculture and Fisheries. We are indebted to Wen-Hsi Yang, at the Centre for Applications in Natural Resource Mathematics of the University of Queensland (Australia), for his invaluable contributions to model formulation.

**Captions to Figs**

Fig. 1.  Historical (2004–2015) variation among catches (t) and fishing effort (boat-days) for the lampara-net fishery targeting eastern sea garfish, *Hyporhamphus australis* off south eastern Australia.

Fig. 2.  Estimated fishing and natural mortalities (± 2 × SD; vertical lines or shaded area) for sea garfish, *Hyporhamphus australis* off south eastern Australia between 2004 and 2015.

Fig. 3.  Estimated (a) recruitment and (b) biomass (± 95% confidence intervals represented by the vertical lines) for sea garfish, *Hyporhamphus australis* off south eastern Australia between 2004 and 2015.

Table 1. Results of fitted mortality models (ranked by increasing value of log-likelihood) estimating the natural mortality ($M$) catchability and selectivity (all ± SE) for eastern sea garfish, *Hyporhamphus australis* off south eastern Australia between 2004 and 2015.

| Model | Log-likelihood | $M$ | Catchability | Selectivity |
|---|---|---|---|---|
| 2 | 3074.7 | $0.52 \pm 0.06$ | $(1.88 \pm 0.15) \times 10^{-3}$ | $s_1 = 0.10 \pm 0.01, s_1 = 0.03 \pm 0.01$ |
| 1 | 3105.0 | $0.43 \pm 0.06$ | $(2.17 \pm 0.15) \times 10^{-3}$ | $s = 0.09 \pm 0.01$ |
| 3 | 3116.6 | Fixed at 0.7 | $(1.58 \pm 0.08) \times 10^{-3}$ | $s = 0.077 \pm 0.004$ |

Table 2.  Estimates of annual fishing mortality (*F*) at age, based on mortality model 2 (which estimated a vector of four parameters; natural mortality (*M*), catchability and a change in selectivity from age 0–1 years onwards) for eastern sea garfish, *Hyporhamphus australis* off south eastern Australia between 2004 and 2015.

| Years | Age group (years) | | | | | |
|-------|------|------|------|------|------|------|
|       | 0–1  | 1–2  | 2–3  | 3–4  | 4–5  | 5–6  |
| 2004–05 | 0.17 | 1.70 | 1.70 | 1.70 | 1.70 | 1.70 |
| 2005–06 | 0.17 | 1.66 | 1.66 | 1.66 | 1.66 | 1.66 |
| 2006–07 | 0.15 | 1.44 | 1.44 | 1.44 | 1.44 | 1.44 |
| 2007–08 | 0.12 | 1.13 | 1.13 | 1.13 | 1.13 | 1.13 |
| 2008–09 | 0.10 | 1.00 | 1.00 | 1.00 | 1.00 | 1.00 |
| 2009–10 | 0.06 | 0.59 | 0.59 | 0.59 | 0.59 | 0.59 |
| 2010–11 | 0.02 | 0.45 | 0.45 | 0.45 | 0.45 | 0.45 |
| 2011–12 | 0.01 | 0.43 | 0.43 | 0.43 | 0.43 | 0.43 |
| 2012–13 | 0.01 | 0.40 | 0.40 | 0.40 | 0.40 | 0.40 |
| 2013–14 | 0.01 | 0.39 | 0.39 | 0.39 | 0.39 | 0.39 |
| 2014–15 | 0.01 | 0.38 | 0.38 | 0.38 | 0.38 | 0.38 |

**Captions to Figs**

Fig. 1.  Historical (2004–2015) variation among catches (t) and fishing effort (boat-days) for the lampara-net fishery targeting eastern sea garfish, *Hyporhamphus australis* off south eastern Australia.

Fig. 2.  Estimated fishing and natural mortalities ($\pm 2 \times$ SD; vertical lines or shaded area) for sea garfish, *Hyporhamphus australis* off south eastern Australia between 2004 and 2015.

Fig. 3.  Estimated (a) recruitment and (b) biomass (95% confidence intervals represented by the vertical lines) for sea garfish, *Hyporhamphus australis* off south eastern Australia between 2004 and 2015.

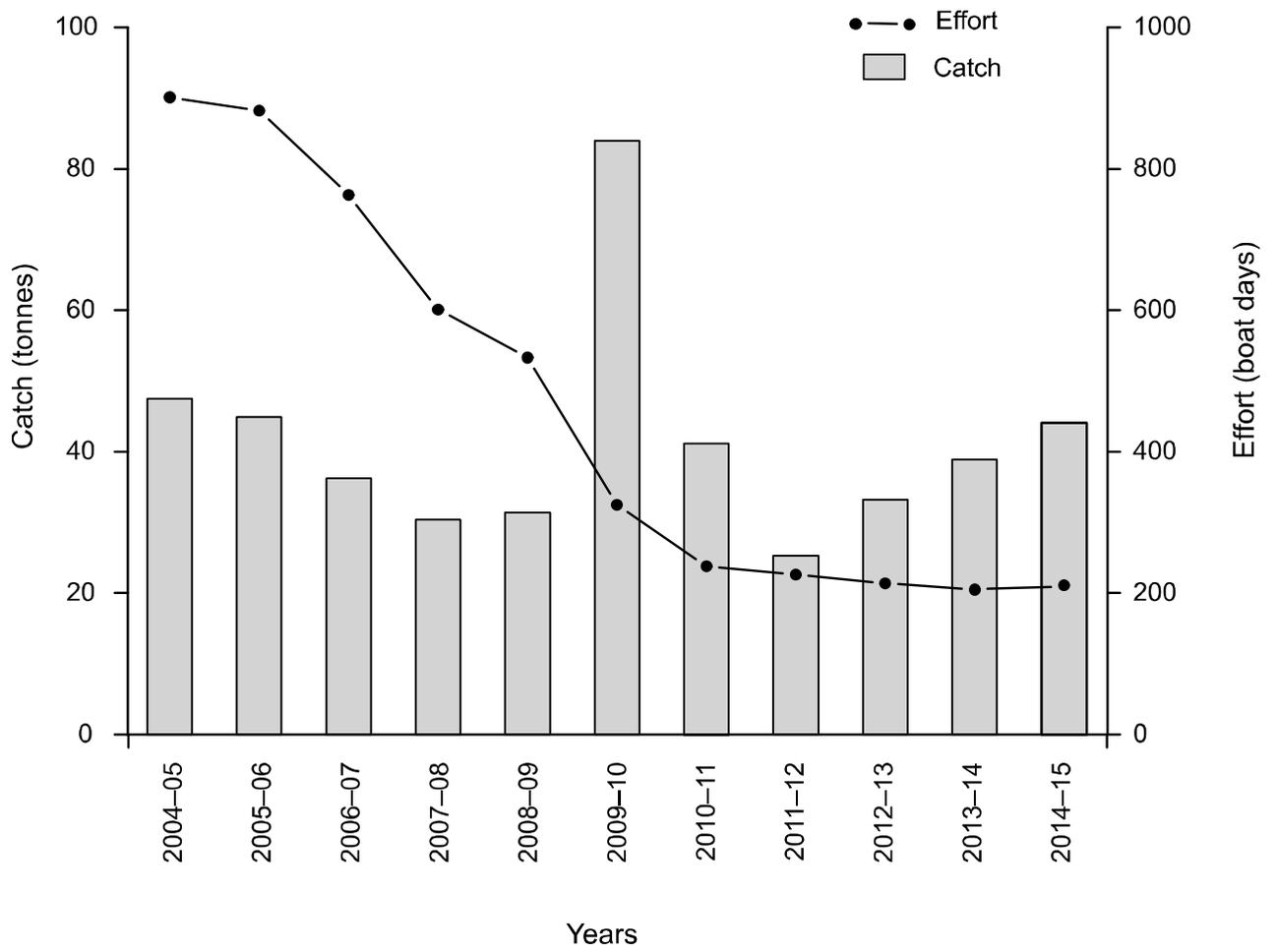

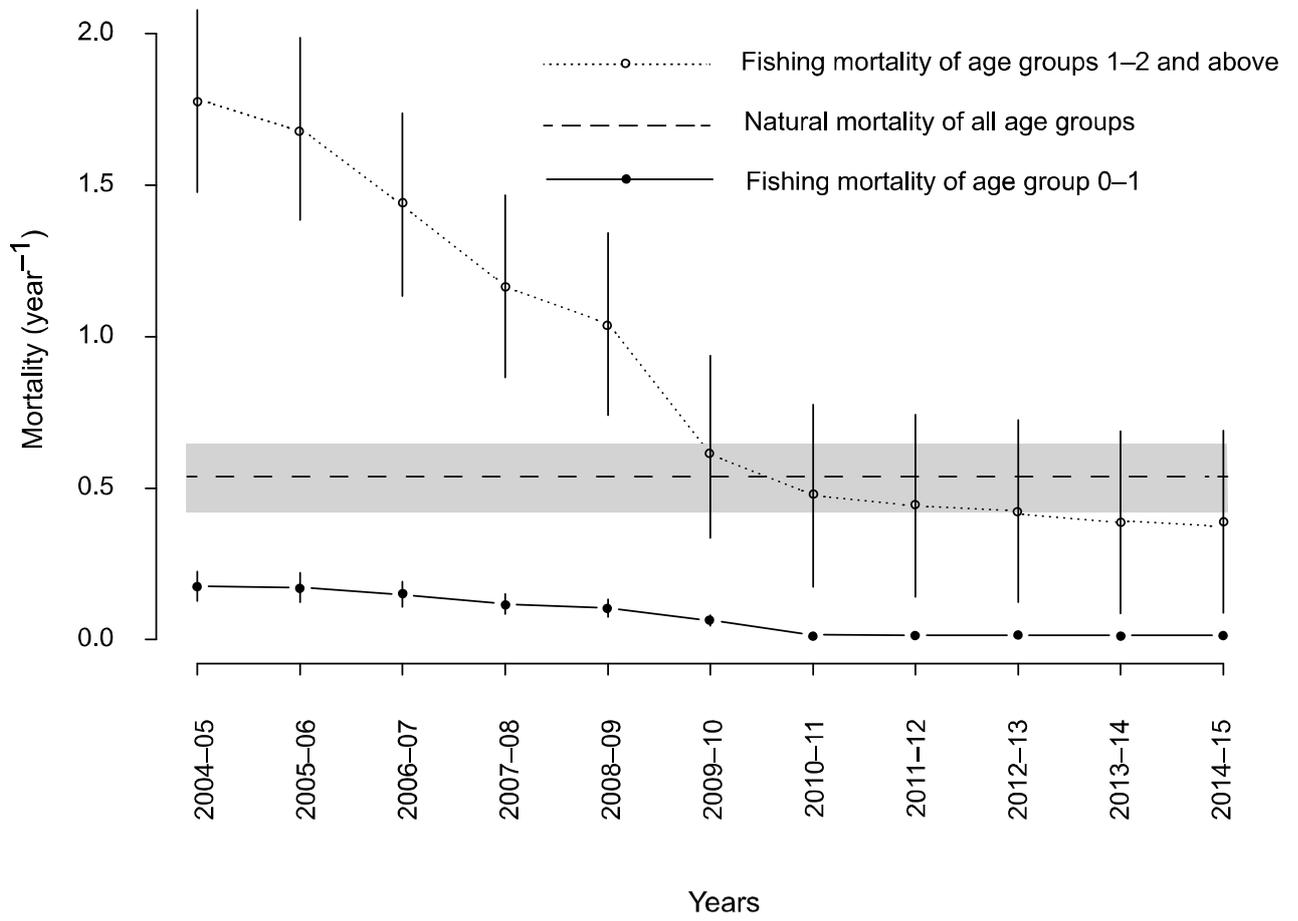

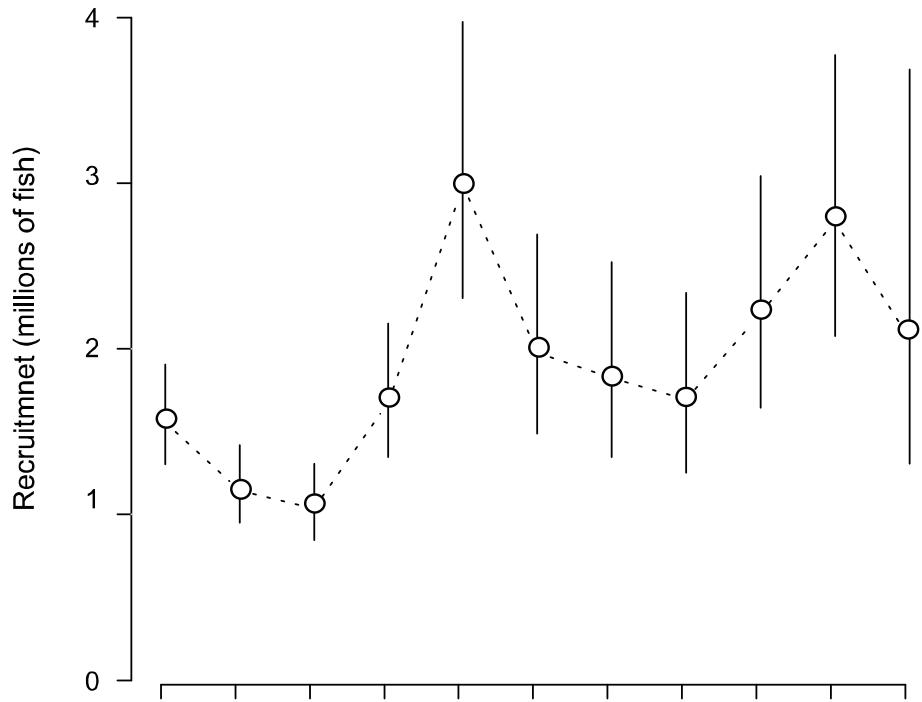

(a) Estimated recruitment of *Hyporhamphus australis*

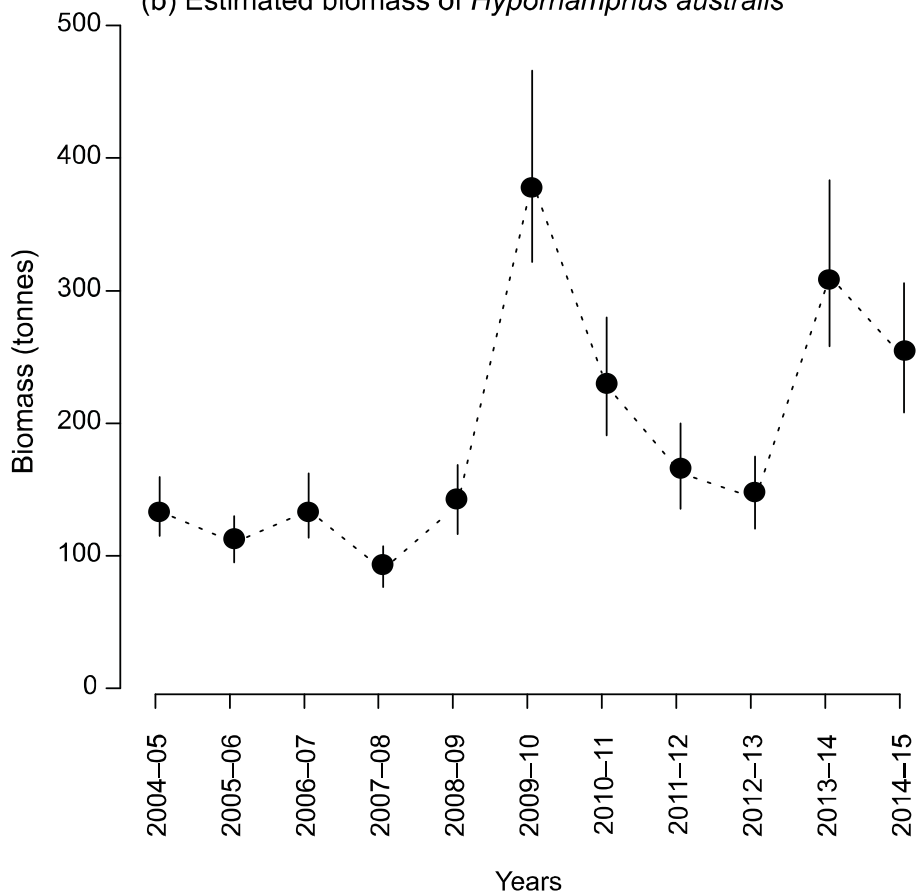

(b) Estimated biomass of *Hyporhamphus australis*

Years

## Supplementary material

Table 1.  Matrix of age samples ($S$) between 2004–05 and 2014–15 for eastern sea garfish, *Hyporhamphus australis* off south eastern Australia.

| | | | Age group (years) | | | |
|---|---|---|---|---|---|---|
| Years | 0–1 | 1–2 | 2–3 | 3–4 | 4–5 | 5–6 |
| 2004–05 | 59.7 | 149.8 | 6.5 | 0.0 | 0.0 | 0.0 |
| 2005–06 | 51.6 | 193.8 | 8.6 | 2.0 | 0.0 | 0.0 |
| 2006–07 | 102.2 | 184.4 | 13.3 | 0.0 | 0.0 | 0.0 |
| 2007–08 | 86.6 | 276.8 | 85.6 | 11.1 | 1.0 | 1.0 |
| 2008–09 | 103.4 | 248.4 | 28.2 | 2.3 | 0.0 | 0.0 |
| 2009–10 | 49.3 | 214.0 | 89.4 | 3.3 | 0.9 | 0.0 |
| 2010–11 | 13.2 | 154.7 | 95.5 | 1.7 | 0.0 | 0.0 |
| 2011–12 | 12.2 | 121.8 | 63.7 | 3.8 | 0.3 | 0.0 |
| 2012–13 | 2.9 | 104.0 | 138.8 | 15.5 | 0.0 | 0.0 |
| 2013–14 | 28.9 | 185.7 | 148.6 | 12.2 | 0.3 | 0.0 |
| 2014–15 | 8.0 | 150.7 | 78.3 | 1.3 | 1.6 | 0.0 |

Table 2. Matrix of weighted age samples ($S^*$) between 2004–05 and 2014–15 for eastern sea garfish, *Hyporhamphus australis* off south eastern Australia.

| | | | Age group (years) | | | |
|---|---|---|---|---|---|---|
| Years | 0–1 | 1–2 | 2–3 | 3–4 | 4–5 | 5–6 |
| 2004–05 | 108.0 | 271.0 | 11.8 | 0.0 | 0.0 | 0.0 |
| 2005–06 | 72.1 | 271.0 | 12.0 | 2.8 | 0.0 | 0.0 |
| 2006–07 | 93.7 | 169.1 | 12.2 | 0.0 | 0.0 | 0.0 |
| 2007–08 | 36.9 | 118.1 | 36.5 | 4.7 | 0.4 | 0.4 |
| 2008–09 | 67.8 | 162.9 | 18.5 | 1.5 | 0.0 | 0.0 |
| 2009–10 | 83.1 | 360.7 | 150.7 | 5.5 | 1.5 | 0.0 |
| 2010–11 | 12.6 | 148.2 | 91.5 | 1.6 | 0.0 | 0.0 |
| 2011–12 | 10.2 | 101.2 | 52.9 | 3.1 | 0.3 | 0.0 |
| 2012–13 | 2.4 | 88.9 | 118.6 | 13.3 | 0.0 | 0.0 |
| 2013–14 | 21.6 | 138.7 | 111.0 | 9.1 | 0.2 | 0.0 |
| 2014–15 | 10.6 | 200.0 | 103.9 | 1.7 | 2.2 | 0.0 |

Table 3. Vectors of yield ($Y$) and effort ($E$) between 2004–05 and 2014–15 for eastern sea garfish, *Hyporhamphus australis* off south eastern Australia.

| Years | Yield (kg) | Effort (boat-day) |
|---|---|---|
| 2004–05 | 47489 | 901 |
| 2005–06 | 44878 | 882 |
| 2006–07 | 36226 | 763 |
| 2007–08 | 30372 | 601 |
| 2008–09 | 31365 | 533 |
| 2009–10 | 84536 | 312 |
| 2010–11 | 40998 | 237 |
| 2011–12 | 25315 | 226 |
| 2012–13 | 33574 | 214 |
| 2013–14 | 38929 | 205 |
| 2014–15 | 43261 | 203 |

Table 4. Matrix of weight at age ($W$) in kg between 2004–05 and 2014–15 for eastern sea garfish, *Hyporhamphus australis* off south eastern Australia.

| Years | Age group (years) | | | | | |
|---|---|---|---|---|---|---|
| | 0–1 | 1–2 | 2–3 | 3–4 | 4–5 | 5–6 |
| 2004–05 | 0.048 | 0.066 | 0.097 | 0.129 | 0.160 | 0.163 |
| 2005–06 | 0.048 | 0.066 | 0.097 | 0.129 | 0.160 | 0.163 |
| 2006–07 | 0.055 | 0.073 | 0.086 | 0.129 | 0.160 | 0.163 |
| 2007–08 | 0.051 | 0.078 | 0.100 | 0.135 | 0.160 | 0.163 |
| 2008–09 | 0.053 | 0.065 | 0.089 | 0.139 | 0.160 | 0.163 |
| 2009–10 | 0.047 | 0.069 | 0.089 | 0.133 | 0.160 | 0.163 |
| 2010–11 | 0.043 | 0.069 | 0.110 | 0.129 | 0.160 | 0.163 |
| 2011–12 | 0.042 | 0.065 | 0.103 | 0.133 | 0.160 | 0.163 |
| 2012–13 | 0.040 | 0.061 | 0.086 | 0.106 | 0.160 | 0.163 |
| 2013–14 | 0.040 | 0.061 | 0.086 | 0.106 | 0.160 | 0.163 |
| 2014–15 | 0.040 | 0.061 | 0.086 | 0.106 | 0.160 | 0.163 |